\documentclass[prl,aps,twocolumn,showpacs]{revtex4}

\usepackage{graphics}

\begin{document}
\title{Qubit versus bit for measuring an integral of a classical field}

\author{ Lev Vaidman and Zion Mitrani}


\affiliation 
{ School of Physics and Astronomy\\
Raymond and Beverly Sackler Faculty of Exact Sciences\\
Tel-Aviv University, Tel-Aviv 69978, Israel}

\date{}

\vspace{.4cm}
 
\begin{abstract}
  Methods for measuring an integral of a classical field via local
  interaction of classical bits or local interaction of qubits passing
  through the field one at a time are analyzed. A quantum method, which
  has an exponentially better precision than any classical method
  we could see, is described.
\end{abstract}


\maketitle Although the retrievable information content of a number of
qubits is essentially equal to the information content which can be
stored in the same number of bits \cite{Ho}, qubits are more efficient
for many specifically tailored tasks.  Recently,  Galv${\rm \tilde a}$o and Hardy
\cite{GH} (GH) found one  such task: pinpointing a particular property
of an integral of a classical field is possible using a single qubit,
but impossible with a classical bit, or even with many bits. This
motivated the current work in which we 
found that qubits are more efficient than bits for a more general
task, namely, the
 measurement of the integral itself.

The task is to measure the integral  of a classical field,
\begin{equation}
\label{eq:int}
I=\int_A^B  \phi (x)  dx,
\end{equation}
via local interactions along the path from $A$ to $B$.  In one case
$N$ bits go one at a time through the path interacting with field, and
in the other case, $N$ qubits pass through the field instead.  In this
task there are no constrains on the complexity of local interactions
between the bits (qubits) and the field. The limitations which we
investigate are due to
the complexity of the carrier
of the information about the field which goes along the path.

In order to make the comparison between quantum and classical methods
simpler, we consider a non-negative classical field and assume that it
is known that the order of magnitude of $I$ is $M$.
The classical method is as follows. The bit  starts at $A$ in the
state 0 and it flips with a probability  proportional to the strength
of the field. Once flipped  to  1,  it
cannot flip back. The  probability  of a flip along the path
is
\begin{equation}
\label{eq:prob}
p =1 - e^{-\lambda I},
\end{equation}
where $\lambda$ has to be optimized for getting the best precision.\cite{foot1}
The exact optimization depends on our a priory information about the probability
for different values of $I$ and the choice of the particular aspect of the precision of the measurement we wish to optimize.
 The precision of measuring $p$ achieved after sending $N$ bits can be
estimated as $\Delta p = {{\sqrt {p(1-p)}}\over \sqrt N}$.  Thus,
essentially for all reasonable approaches, the uncertainty of measuring
$I$ is of the order of $M\over \sqrt N$.\cite{foot2}

 In the quantum method, we arrange an interaction which
leads to a qubit ``rotation''   proportional to the
strength of the field, for example, a  spin precession 
in  a  magnetic field. The method works equally well  for
negative  and  general  classical  fields.  The   only
important  parameter is the range of possible  values  of
the integral. The strength of the coupling should be such
that  for  every   value  of the  integral  we  will  get
different  final state of the qubit. Thus, we have to choose the strength
of the coupling to the field such that a rotation of more than  $2\pi$ will  not be probable.

When we send $N$ spin-$1\over 2$ particles  through the field, they all rotate by the same
angle. If we start with all spins pointing in, say $x$ direction, and
the magnetic field in the  $z$ axis, the direction of the
spin in the $x-y$ plane will yield the value of the integral.

The precision of measuring direction of $N$ parallel spins is
proportional to $N^{-{1\over 2}}$. This is the same dependence as in
the probabilistic methods with bits.  However, Peres and Scudo
\cite{PS} (see also Bagan et al. \cite{Ba}) showed that taking $N$
entangled spins one can reduce the uncertainty to be proportional to
$N^{-1}$. They  optimized  defining the direction in
three dimensions with $N$ spin-$1\over 2$ particles, while in the
present problem we need to find direction only in two dimensions. So,
some further optimization is possible, but  the uncertainty remains
proportional to $N^{-1}$. This concludes the description of our 
quantum method I for measurement of  integral $I$.

A crucial requirement for the advantage of qubits versus bits is
that we are only allowed to send bits and qubits one at a time. If we send
all bits together, we can build a probabilistic counter which can register up to
$2^N$ counts. The counter makes counts while moving in
the field with the probability for a count proportional to the field. The uncertainty
in the total number of counts is of the order of the square root of
this number. Therefore, adjusting
the strength of the interaction such that the expectation value of the
number of counts is of the order of $2^N$,
we can get the
uncertainty in the measurement
of $I$ proportional to $2^{-{N\over 2}}$. The uncertainty is
exponentially smaller than in the quantum method described above.

Of course, if qubits are allowed to pass along the field together, the
precision that can be achieved in the measurement of the integral is
not smaller than that in the classical case. But can exponentially small
uncertainty be achieved with qubits passing through the field one at a
time? As we will show below, the answer is yes, if we allow various strengths of the
interaction between the qubits and the field.

 On the other hand, we do
not see how this freedom  might  lead to a significant advantage in the
classical case. Indeed, it is intuitively clear (especially after the
GH proof for a particular case) that a bit passing through the field
cannot have a deterministic information about the integral of the
field. Then, the only way we can imagine for storing  information  in the bit is
in the value of its probability  to be 1. When we are
given $N$ bits, the
uncertainty in the measurement of the probability decreases as
$N^{-{1\over 2}}$. It seems to us highly implausible that there is a classical
method which do better than this.

Now we turn to the description of the method (referred hereafter as
quantum method II) which employs qubits interacting with various
strengths with the field. The strengths depend only on the number of
the qubit passing through the field and are fixed before the
experiment.
No
additional structure is required for the qubit.

 The basis of this method is
the result of  Galv${\rm \tilde a}$o and Hardy \cite{GH}.  They considered a
particular case in which
\begin{equation}
\label{eq:intm}
I =m\alpha,
\end{equation}
where $\alpha$ is known and $m$ is an integer. GH found a method 
which allows  with a single qubit to answer the question: is $m$ even or odd?\cite{foot3} 
 They achieved the goal by
tuning the strength of the interaction in such a way that $I= \alpha$
yields a rotation by $\pi$. Thus,  for an odd integer the spin flips, and
for an
even integer, it returns to its initial state.

 If we send a number
of qubits, one after the other, we can modify this procedure to find $m$ itself. To this end it
is arranged that the qubits we send interact with the field with
different strengths: the first as in the GH protocol, the second with
half of the strength of the first, the next with the half of the
strength of the preceding, etc. In the first step we find the last
digit of $m$ written in the binary way. In the next step we find the
preceding digit, and thus, after $k$'th step we find $m({\rm mod} 2^k)$.

The procedure works in the following way. If the last digit is zero, then in the second step we  repeat the
GH protocol with half of the strength of the interaction. Since now
we know that $I =m'2\alpha$, the protocol determines  the last digit of 
$m'$ which is second digit from the end of $m$. If the last  digit of $m$ is $1$, in the second step we
should modify the 
procedure by additional rotation of  the spin
 by the angle $\theta_2 = -{\pi \over 2}$. In the $k$'th step we should
compensate for all non-zero digits, as follows:
\begin{equation}
\label{eq:corr}
\theta_k =- \sum_{i=1}^{k-1}{{\pi d(i)} \over 2^{(k-i)}}.
\end{equation}
where $d(i)$ is the value of digit number $i$ from the end. The
method  yields  one digit   for each qubit  and  yields
zeros  once   the whole  number  is  written.

In a general case, when our only prior  information about $I$ is its
order of magnitude, we can combine the two methods we described
above. We chose $\alpha$ and, using  method I, we find the reminder $\beta$ of
the division of $I$ by $\alpha$. Then, using  method II, we find $m$:
\begin{equation}
\label{eq:intmm}
I =m\alpha +\beta .
\end{equation}

After measuring $\beta$ with a good precision we  apply 
method II for finding $m$. The reminder $\beta$ requires
additional correction angles in the application of the second method.
 For the first step, the correction angle is $\theta_1=
-{{ \pi \beta}\over \alpha}$ and, in general,  for   $k$'th step, the correction
angle is:
\begin{equation}
\label{eq:corr1}
\theta_k =- \sum_{i=1}^{k-1}{{\pi d(i)} \over 2^{(k-i)}} -{{ \pi \beta}\over{2^{k-1} \alpha}}.
\end{equation}

The requirement for choosing
$\alpha$ is that we will have enough qubits to find all digits of $m$.
We get high probability for that if
\begin{equation}
\label{eq:ineq}
\alpha = {{10 M} \over 2^{N-N_0}}  ,
\end{equation}
 where $N_0$ is the number of qubits
used in the measurement of $\beta$. 
If the probability of error in the measurement of $m$ is negligible (when
$\beta$ is measured with the high precision), then the uncertainty in
the value of $I$ is, essentially, the uncertainty in the measurement
of $\beta$ which, in the best case (Peres-Scudo method), is of the
order of ${\alpha \over N_0}$.

 However, as the anonymous referee
pointed out, the error in the measurement of $m$ turns out to be not
too large even if there is a large error in the measurement of
$\beta$. In fact, it is more effective not to ``waste'' qubits on
measurement of $\beta$. The quantum method II works well by itself even
in a general case. Now $\alpha$ becomes much smaller:
\begin{equation}
\label{eq:alpha}
\alpha = {{10 M} \over 2^N}  .
\end{equation}

 In order to estimate the precision of the quantum method we calculate
 probability of a particular reading $m$ given the actual value of
 $I$. We will show that it is rapidly decreasing function of the
 difference between the actual value of the integral $I$ and the read
 out of the device $\tilde I= m\alpha$.
In our procedure, the last digit of $m$ is specified by the spin
measurement. The spin rotates by the angle $\Theta_1 ={{I\pi}\over \alpha}$ and
is found in the direction specified by the angle $\tilde \Theta_1    ={{\tilde I\pi}\over \alpha}$. Thus, the
probability for this particular outcome of the spin measurement which
specifies the last digit is
$p_1=\cos^2{{\Theta_1 -\tilde \Theta_1}\over 2}=\cos^2{{(I-\tilde I)\pi}\over {2 \alpha}}$.  The  digit $k$ from the end, $k>1$, is specified
by the measurement of the spin which is rotated by the angle
\begin{equation}
  \label{eq:teta2}
 \Theta_k = {{I\pi}\over {2^{k-1}\alpha}} + \theta_k
\end{equation}
and found  in the direction specified by the angle 
\begin{equation}
  \label{eq:teta2til}
\tilde \Theta_k ={{\tilde
    I\pi}\over {2^{k-1} \alpha}} +\theta_k.
\end{equation}
 Therefore, the
probability for this particular outcome of the spin measurement which
specifies the $k$'th digit is
$p_k =\cos^2{{(I-\tilde I)\pi}\over {2^k \alpha}}$.
Thus, the probability for readout  $\tilde I$ given the actual value
of the integral $I$ is:
\begin{equation}
  \label{eq:probm}
p(\tilde I|I)=\prod_{k=1}^N{}\cos^2{{(I-\tilde I)\pi}\over {2^k \alpha}}.
\end{equation}
 Denoting the error $\delta I$ and substituting the  value of
$\alpha$ from (\ref{eq:alpha}), the probability of the error 
is
\begin{equation}
  \label{eq:probdelta}
p(\delta I)=\prod_{k=1}^N{}\cos^2{{\delta I \pi}\over {2^{k-N} 10 M}}.
\end{equation}
This function vansihes for $\delta I=n\alpha$ for integer $n$, except
for $n=0$, where it has maxima which is equal 1. It has to be the so, because the method
yields no errors for $ I=m\alpha$.  The function has local maxima at $\delta
I=(n+{1\over 2})\alpha$, except for $n=-1,0$.

 The read-out value of the integral, $\tilde I$, might have only
discrete values, $m\alpha$. Therefore, the error of the
order of $\alpha$ is unavoidable.
 In the worse case, $I{\rm mod} \alpha = {\alpha
  \over 2}$,  but even in this case we have only small probability to get
the error which is an order of magnitude larger than
$\alpha$. For large $N$, this probability has almost  no depndence on  $N$; we calculate it for $N=30$:
\begin{eqnarray}
  \label{eq:probdelta1}
  p(\delta I> 10\alpha )=1- p(\delta I< 10\alpha
  )\nonumber\\= 1- \sum_{n=-9}^{10}\prod_{k=1}^{30}{}\cos^2{{({1\over 2}
      +n)\alpha\pi}\over {2^{k-N} 10
      M}}\nonumber\\=1 -\sum_{n=-9}^{10}\prod_{k=1}^{30}{}\cos^2{{({1\over 2}
      +n)\pi}\over {2^k}} \simeq 0.019.
\end{eqnarray}
Thus, we should expect an error of the order of $\alpha$ (it is of the order of $10^{-7}$ in our case) and not significantly larger.

In order to illustrate our result we performed computer simulation of
classical ($30$ bits) and quantum ($30$ qubits)  measurements of the integral for  ten different fields.
 We took   $I_n=(n\pi){\rm mod}10$, $~n=1,2,...10$.  We assumed that the order of magnitude of the integral is given,
$M=5$, and we choose parameter $\lambda$  of the
classical method such that the precision of the measurement of
$I$ for $I=M$ is optimized. The uncertainty in classical measurement can be estimated as
\begin{equation}
  \label{eq:probdelta1}
\Delta I={{\sqrt {e^{\lambda I} -1}}\over {\lambda \sqrt N}}.
\end{equation}
For  $I=M$ it has a minimum around  $\lambda={ 1.2\over M}$ and for parameter we chose, the uncertainty  $\Delta I$ is of the order of 1.

  The results are shown in
Table 1. We see that the error in classical method is indeed of the order of 1 and the quantum error is of the order of $10^{-7}$. 
\begin{table}\begin{tabular}{|c|c|c|c|}
  \hline
$n$ & $I=n\pi mod(10)$ & quantum & classical \\\hline
  1 & 3.141592654 & 3.141592494 & 3.175583382 \\
  2 & 6.283185307 & 6.283185389 & 4.577585162 \\
  3 & 9.424777961 & 9.424777867 & 9.594107747 \\
  4 & 2.566370614 & 2.566370611 & 1.689440418 \\
  5 & 5.707963268 & 5.707963268 & 5.016553197 \\
  6 & 8.849555922 & 8.849555813 & 9.594105057 \\
  7 & 1.991148575 & 1.991148466 & 2.619198463 \\
  8 & 5.132741229 & 5.132741166 & 8.395441798 \\
  9 & 8.274333882 & 8.274333865 & 6.706033821 \\
  10 & 1.415926536 & 1.415926495 & 0.929766618 \\\hline
\end{tabular}
\caption{The results of simulation of classical and quantum measurements of ten values of $I$,
 $I_n=(n\pi){\rm mod}10$. Quantum method uses 30 qubits and classical method uses 30 bits.}
\end{table}

The technology today is far from getting this exponential advantage of
the quantum method which requires stability and high precision at very
large range of the interaction strength. Also, preparation of initial
entangled states and complicated collective measurements of
Peres-Scudo measurements are difficult for experimental
implementation. However, recent progress in quantum information
experiments allows us to believe in the prospects of at least partial
implementation of our proposal which will manifest advantage of the
quantum method.

It is a pleasure to thank Yakir Aharonov, Shmuel Nussinov, Benni
Reznik, Boris Tsirelson, and Stephen Wiesner for helpful discussions. This research was supported in part
by grant 62/01 of the Israel Science Foundation and by the Israel MOD
Research and Technology Unit.




\begin{thebibliography}{9}
\bibitem{foot1}
 This method and the quantum method, which is described below, are
also applicable for  measuring the sum of the field values at discrete
points, $\sum_i  \phi (x_i)$. In this case, the interaction at each
point should be such that the probability of the flip (if it has  not been
flipped already) is
$
p_i=
1-e^{-\lambda \phi(x_i)}.
$
\bibitem{foot2}
The method has to be modified if the field is allowed to be negative
as well as positive. In this case, we can send half of the bits for
measuring the positive contribution of the field and half of the bits for
measuring the negative contribution of the field. (An alternative method,
suggested by the referee, is
to add to the field a constant which will make all values positive.)  This, of course, 
decreases the precision (especially  when the positive and negative
contributions are much larger than $M$), but the overall dependence of
the uncertainty on
$N$  remains (for large $N$) proportional to $N^{-{1\over 2}}$.
\bibitem{foot3} Note that our classical probabilistic method allows to
  answer the GH question with an almost vanishing probability of an error
  when we are given $N$ bits, such that ${\sqrt N} \gg M$, where $M$
 (assumed to be known) is the estimation of the bound for $m$.
\end{thebibliography}
\end{document}